\documentclass[fleqn,usenatbib]{mnras}

\usepackage{newtxtext,newtxmath}

\usepackage[T1]{fontenc}

\DeclareRobustCommand{\VAN}[3]{#2}
\let\VANthebibliography\thebibliography
\def\thebibliography{\DeclareRobustCommand{\VAN}[3]{##3}\VANthebibliography}

\usepackage{graphicx}	
\usepackage{amsmath}	
\usepackage{pdflscape}
\usepackage{afterpage}
\usepackage{orcidlink}

\newcommand{\athena}{{\it Athena}}

\title[AGN X-ray Spectroscopy with Neural Networks]{AGN X-ray Spectroscopy with Neural Networks}

\author[M. L. Parker et al.]{M. L. Parker,$^{1}$\thanks{E-mail: mlparker@ast.cam.ac.uk}\orcidlink{0000-0002-8466-7317}
M. Lieu,$^{2}$\orcidlink{0000-0002-4487-8136}
G. A. Matzeu,$^{3,4}$\orcidlink{0000-0003-1994-5322}
\\
$^{1}$Institute of Astronomy, University of Cambridge, Madingley Road, Cambridge, CB3 0HA, UK\\
$^{2}$ School of Physics $\&$ Astronomy, University of Nottingham, Nottingham, NG7 2RD\\
$^3$Department of Physics and Astronomy (DIFA), University of Bologna, Via Gobetti, 93/2, I-40129 Bologna, Italy\\
$^{4}$INAF-Osservatorio di Astrofisica e Scienza dello Spazio di Bologna, Via Gobetti, 93/3, I-40129 Bologna, Italy\\
}

\date{Accepted XXX. Received YYY; in original form ZZZ}

\pubyear{2022}

\begin{document}
\label{firstpage}
\pagerange{\pageref{firstpage}--\pageref{lastpage}}
\maketitle
\begin{abstract}
We explore the possibility of using machine learning to estimate physical parameters directly from AGN X-ray spectra without needing computationally expensive spectral fitting. Specifically, we consider survey quality data, rather than long pointed observations, to ensure that this approach works in the regime where it is most likely to be applied. We simulate \athena\ WFI spectra of AGN with warm absorbers, and train simple neural networks to estimate the ionisation and column density of the absorbers. We find that this approach can give comparable accuracy to spectral fitting, without the risk of outliers caused by the fit sticking in a false minimum, and with an improvement of around three orders of magnitude in speed.
We also demonstrate that using principal component analysis to reduce the dimensionality of the data prior to inputting it into the neural net can significantly increase the accuracy of the parameter estimation for negligible computational cost, while also allowing a simpler network architecture to be used.
\end{abstract}

\begin{keywords}
galaxies: active -- accretion, accretion disks -- techniques: spectroscopic -- black hole physics 
\end{keywords}


\section{Introduction}
\label{sec:intro}
The fundamental aim of X-ray spectroscopy is to estimate physical parameters. For AGN, this is typically concentrated on a few key parameters such as black hole spin \citep[e.g.][]{Risaliti13}, outflow velocity \citep[e.g.][]{Nardini15}, or absorbing column \citep[e.g.][]{Kaastra14}. The prevailing paradigm for this parameter estimation is spectral fitting, where a physically motivated spectral model is constructed and its parameters adjusted until it provides a good match to the data, usually with a spectral fitting package like \textsc{xspec} \citep[][]{Arnaud96}. This is arguably the most intuitive approach to take to the problem, and has been very effective over the lifetime of X-ray astronomy.

There are some drawbacks with this approach, however. Spectral fitting like this can rapidly become very computationally expensive, and is very difficult to automate without heavy supervision, as models can frequently achieve good fits to the data that do not correspond to physically meaningful parameter combinations. As the size of astronomical datasets expands with the next generation of instrumentation (such as \athena ), the computational cost of running spectral fits on huge numbers of spectra is likely to become prohibitive. This is likely to be particularly acute for AGN and other compact objects, as their time variability means that their parameters can change very rapidly and high cadence time-resolved spectroscopy is both possible and highly desirable.

Machine learning offers some alternative tools which can be used for parameter estimation, which can in principle be applied to the problem of X-ray spectroscopy. While generally less intuitive, these methods can scale very effectively to large datasets, drastically reducing the computational expense involved. Artificial neural networks (ANN), designed to mimic how biological neurons function \citep{Mcculloch1943logical}, can learn to predict an arbitrary set of outputs, such as physical parameters, from an arbitrary set of inputs, such as spectral count rates. If they can be trained to reliably predict parameters of interest then we can potentially use neural networks to replace spectral fitting \citep[e.g.][]{firth2003estimating, vanzella2004photometric} for large datasets that would otherwise take an unreasonable amount of time to model. {Most relevant here, \citet[][]{Ichinohe18} showed that a neural network can in principle be used to estimate parameters from simple X-ray microcalorimeter spectra of clusters.}

In this work, we explore how a machine learning approach can be used to estimate parameters from survey quality data, with limited energy resolution and short exposures. We use a combination of neural networks and spectral decomposition tools to estimate the parameters of a warm absorber in \athena\ Wide Field Imager \citep{Meidinger2017wide} (WFI; \athena 's survey instrument) spectra. Warm absorbers are detectable even in low signal spectra \citep[e.g.][]{Reynolds97}, so are likely to be detected in large numbers by \athena , potentially requiring large-scale automated parameter estimation for population studies. 
{The spectra in this regime are both noisier and lower resolution than those considered by previous authors, and we include the effects of multiple independent spectral components, making the spectra significantly more complex. We also explore the impact of pre-processing the data with principal component analysis to reduce the dimensionality, a powerful technique that can dramatically improve the performance of neural networks.}
In section~\ref{sec:data} we simulate a set of spectra for training and testing. In section~\ref{sec:results} we consider three different methods for recovering the parameters of the warm absorber in the simulated spectra: automated spectral fitting, a neural network, and a neural net with PCA pre-processing. In section~\ref{sec:discussion} we explore the implications of this work, in particular how neural networks could be used on a larger scale with the next generation of X-ray telescopes such as \textit{XRISM} and \textit{ATHENA}. Finally, we summarise our conclusions in section~\ref{sec:conclusions}.

\section{Simulated Data}
\label{sec:data}
In a similar manner to \citet{Parker22}, we set up a physically motivated model in \textsc{xspec} version 12.12.0, then use the \textsc{pyxspec} python wrapper \citep{Gordon2021pyxspec} to simulate a large number of spectra. We use a relatively simple model of a power-law continuum, a phenomenological soft excess, Galactic absorption \citep[modeled with \textsc{tbabs}][]{Wilms00} and a single layer of warm absorption \citep[modelled with an \textsc{xspec} table model version of the \textsc{xabs} model from \textsc{spex}][]{Parker19_mrk335, Steenbrugge03, Kaastra96}. The full model, in \textsc{xspec} format, is \textsc{tbabs $\times$ xabs $\times$ (blackbody $+$ powerlaw)}.

\begin{figure}
    \centering
    \includegraphics[width=\linewidth]{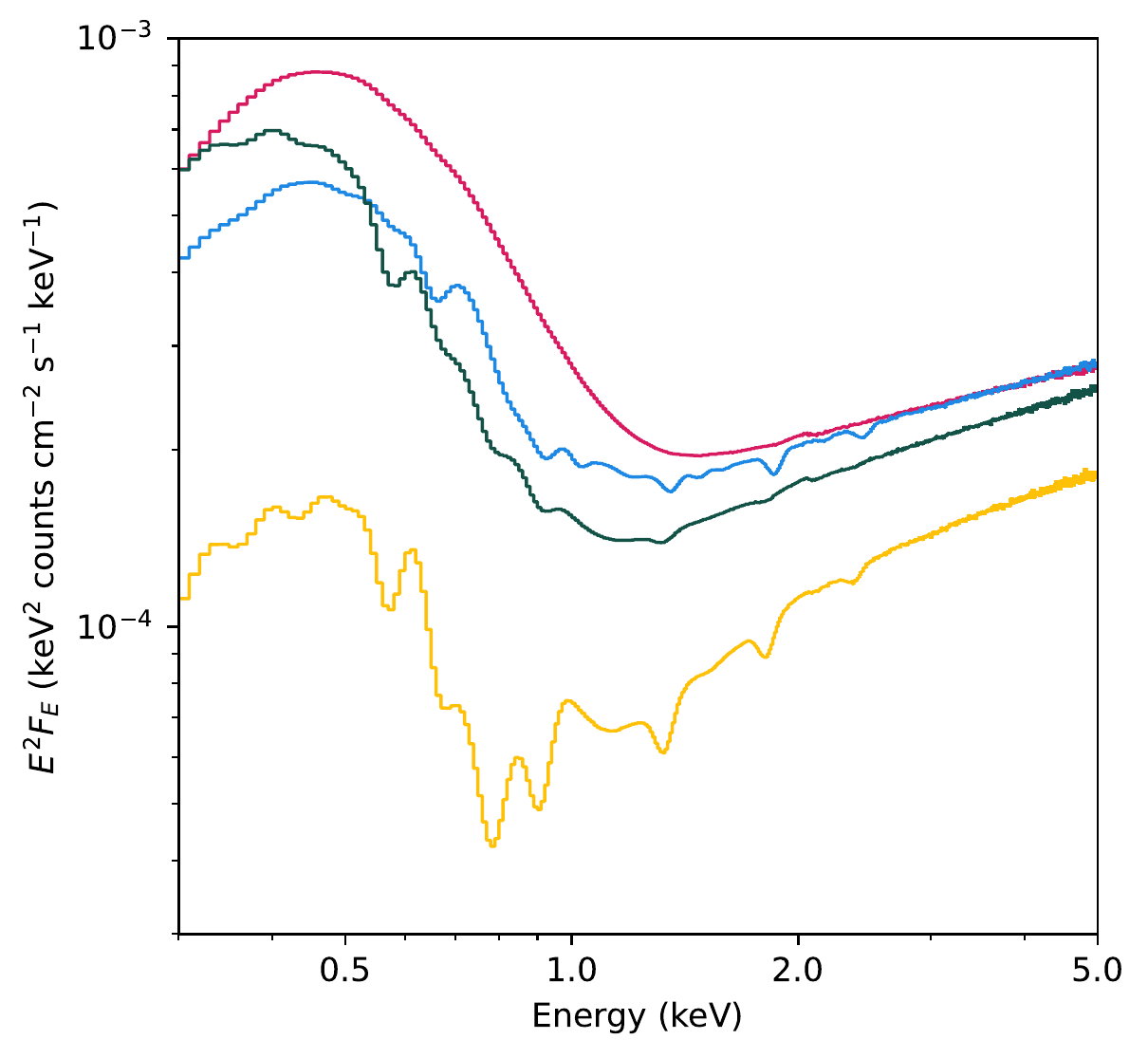}
    \caption{4 randomly selected simulated \athena\ WFI spectra from the sample, showing the warm absorber, black body soft excess and powerlaw continuum.}
    \label{fig:randomspectra}
\end{figure}

We draw 11000 parameter combinations (details of the distributions used for each parameter are given in Table~\ref{tab:param_dists}), and simulate a 10~ks WFI spectrum using the \textsc{xspec} \texttt{fakeit} command. The first 10000 spectra are to be used as the training set for the neural network, and the last 1000 as the test and validation set and for comparison with automated fitting in \textsc{pyxspec}. The training set is used exclusively to train the network, while the test/validation set is used to evaluate the performance of the network on data that it has not been trained on.
We show four randomly selected spectra in Fig.~\ref{fig:randomspectra}.

\begin{table*}
    \centering
    \caption{Parameter distributions used to generate the synthetic spectra. The warm absorber parameter ranges are roughly based on the range of ionisations and column densities found by \citet[][]{Laha14}.}
    \begin{tabular}{l c c r}
    \hline
    \hline
    Model & Parameter & Distribution & Description/unit\\
    \hline
    \textsc{tbabs}     & $N_\mathrm{H}$ & Log normal, $\mu=10^{20}$, $\sigma=0.1$ dex & Galactic column density (cm$^{-2}$)\\
    \textsc{xabs}   & $\log(\xi)$ & Uniform, 1--3 & Ionization (erg~cm~s$^{-1})$\\
                    & $N_\mathrm{H}$ & Log uniform, $10^{20}$--$10^{22}$ & Warm absorber column density (cm$^{-2}$)\\
    \textsc{blackbody} & $kT$ & Normal, $\mu=0.1$, $\sigma=0.01$ & Temerature (keV)\\
                    & $F_\mathrm{0.5-10}$ & Log normal, $\mu=10^{-12.5}$, $\sigma=0.1$ dex & 0.5--10~keV flux (erg~s$^{-1}$~cm$^{-2}$) \\
    \textsc{powerlaw}   & $\Gamma$ & Normal, $\mu=1.8$, $\sigma=0.1$ & Photon index\\
                    & $F_\mathrm{0.5-10}$ & Log normal, $\mu=10^{-12}$, $\sigma=0.1$ dex &  0.5--10~keV flux (erg~s$^{-1}$~cm$^{-2}$) \\
    \hline
    \end{tabular}
    \label{tab:param_dists}
\end{table*}

\section{Results}
\label{sec:results}
We will now explore three different approaches to estimating parameters from the simulated spectra. We evaluate the methods by their ability to recover the ionisation parameter and column density of the warm absorber, assumed to be the primary scientific interest of this hypothetical study, and the time it takes for them to complete the parameter estimation. We test all three methods on the same laptop, running on an 8 core Intel i9-10885H CPU. 

\subsection{Spectral fitting}
\label{sec:spectralfitting}

Firstly, as a comparison for the machine learning approaches, we use a conventional spectral fitting approach using \textsc{pyxspec} to automate the fitting. Each spectrum is binned to a minimum signal to noise ratio of 6, and fit from 0.3--5~keV with the same model used to simulate the spectra. For each fit, the parameters are re-set to values in the middle of each parameter distribution and the model is renormalised. 

\begin{figure*}
    \centering
    \includegraphics[width=0.8\linewidth]{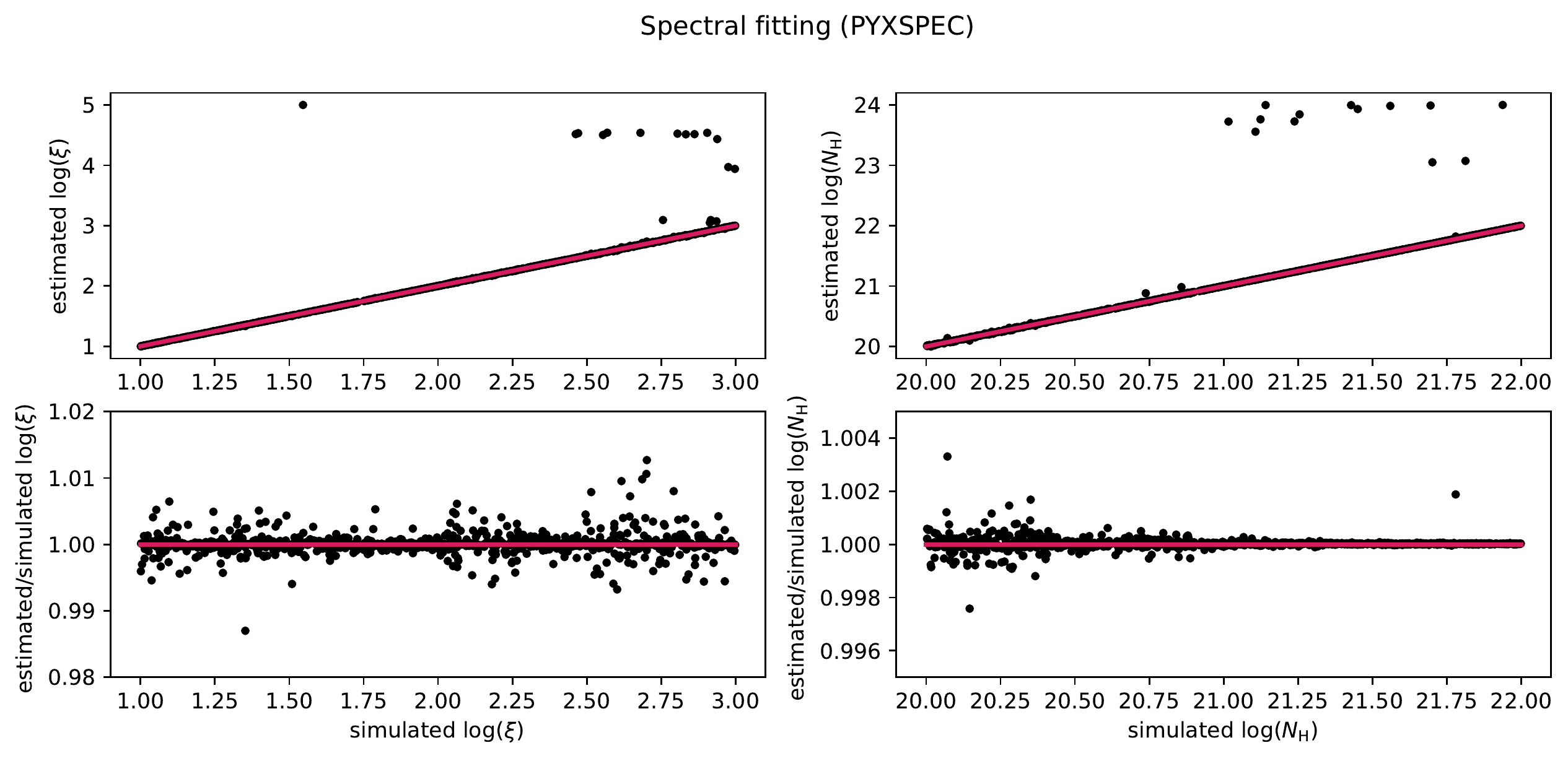}
    \includegraphics[width=0.8\linewidth]{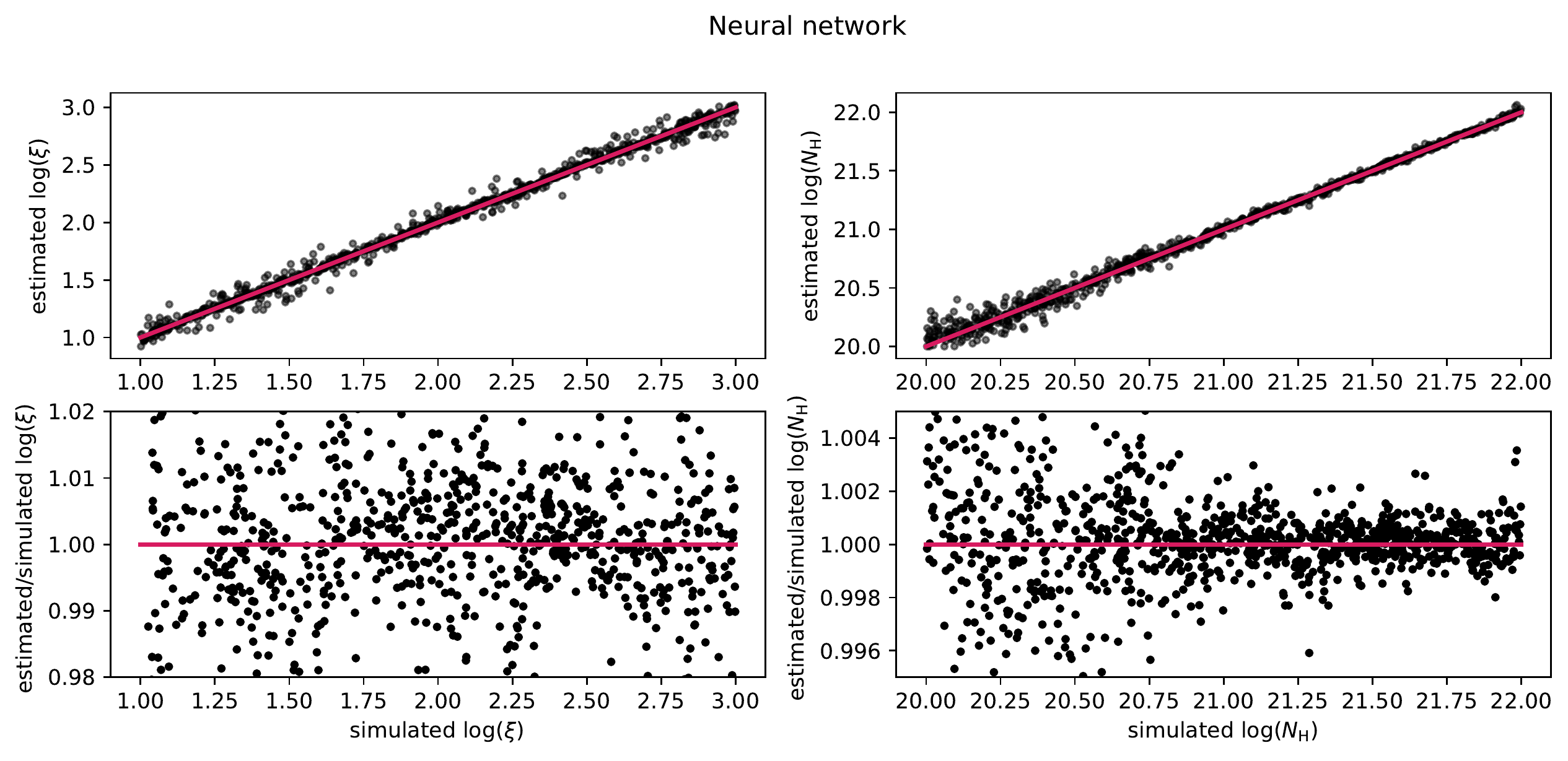}
    \includegraphics[width=0.8\linewidth]{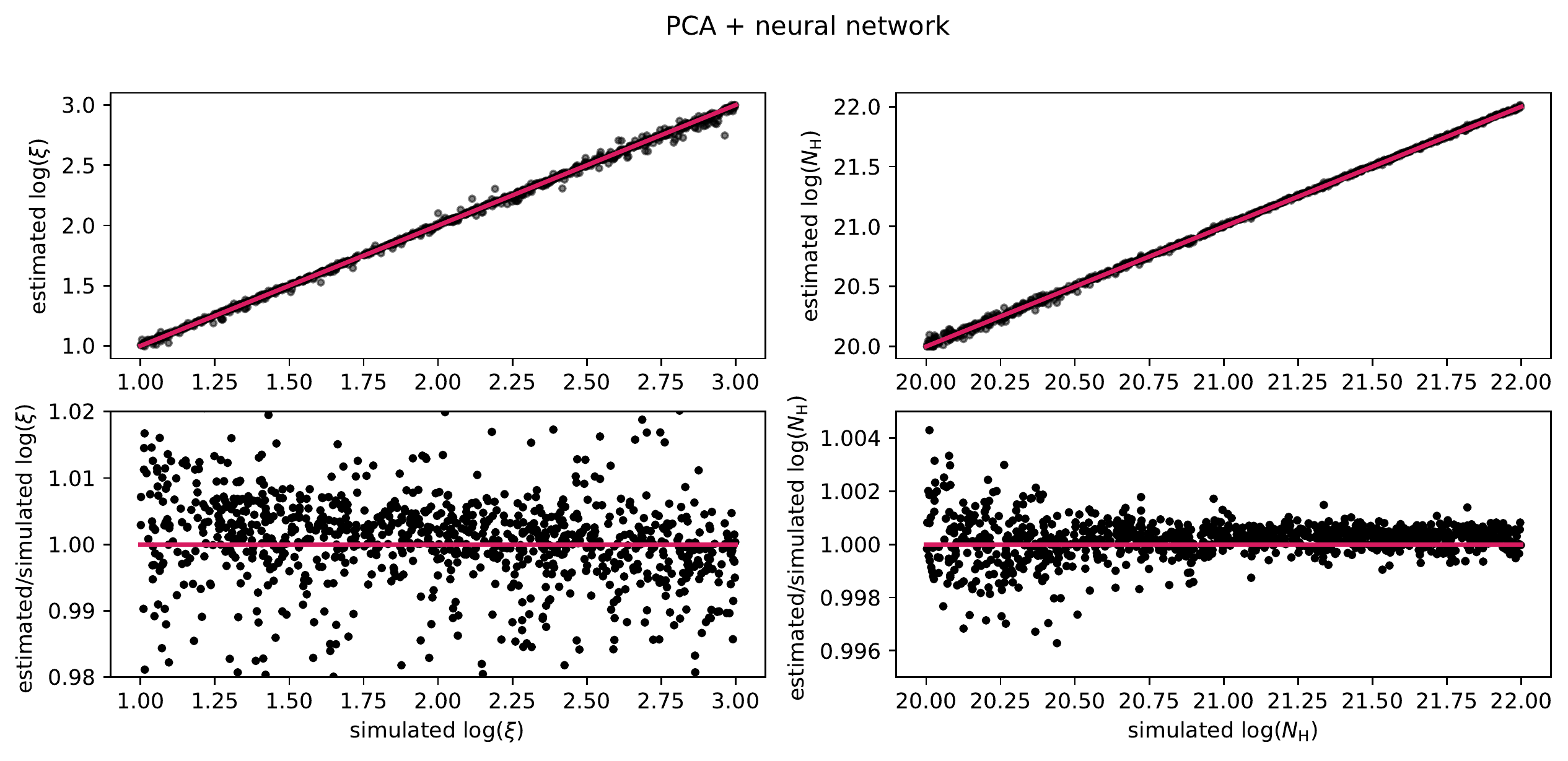}
    \caption{Parameter recovery for each of the three methods. In each case, for the ionisation and column density of the warm absorber, we plot the estimated parameter value as a function of the simulated value in the top panel, and the ratio of the two as a function of simulated value in the bottom panel (zoomed to exclude outliers). The red line shows the 1:1 relation. Conventional spectral fitting is the most accurate when outliers are excluded, however it also has a tendency to get stuck in false minima. The standard neural network performs well, recovering the parameters well but with a small scatter. Including the PCA pre-processing step dramatically increases the accuracy of the neural network, for a negligible computational cost. }
    \label{fig:parameter_estimates}
\end{figure*}

We then use the standard \textsc{xspec} fit algorithm, with no manual oversight, to estimate the parameters by minimising the $\chi^2$ statistic\footnote{see \url{https://heasarc.gsfc.nasa.gov/xanadu/xspec/manual/XSappendixStatistics.html}}. We note that the accuracy of the fits could likely be improved by running the error algorithm as well, as \textsc{xspec} fits frequently get stuck in false minima and the more exhaustive parameter search run by the error estimation algorithm can sometimes escape if the minimum is shallow. However, this would also drastically increase the computational expense, which we are aiming to minimise, and does not help if the false minimum is deep enough.

We show the best-fit values of ionisation and column density, plotted against the simulated values, in the top panels of Fig.~\ref{fig:parameter_estimates}. The parameter estimation is generally extremely good, recovering the corrects value with very small scatter, with the exception of a handful of points that find false minima, preferentially at high column densities and high ionizations. Fitting the 1000 spectra of the testing set, running in parallel on 8 cores, takes 5 minutes 30 seconds. While this is an entirely reasonable run time, we note that we deliberately set up a very simple model for testing purposes, and the computational cost is likely to be much higher in real applications with more complex models and more spectra.

\subsection{Neural net}
\label{sec:neuralnet}

We next consider directly estimating the parameters from the raw spectra using a simple neural network. The network consists of several layers of artificial neurons, starting with an input layer, followed by some number of hidden layers, and finishing with an output layer with one neuron per output parameter. Each neuron takes an input computed from the outputs of the all neurons in the previous layer, and consequently, passes an output to all neurons in the next layer. The output is calculated by an activation function, applied to a weighted sum of the inputs plus a bias term. The training process adjusts the weights and biases of each neuron until the output from the final layer matches the parameters, attempting to minimise a loss function. 

To build the network, we use the \textsc{Keras} deep learning API \citep{keras} for \textsc{python}, running on the \textsc{TensorFlow} machine learning platform \citep[][]{tensorflow}.
We use the \texttt{Adam} optimiser \citep{Kingma14} with a learning rate of $10^{-4}$, the Rectified Linear Unit (ReLU) activation function \citep{Nair2010rectified}, and the mean squared error loss function. We implement early stopping \citep{Prechelt1998early} with a patience of 50 to prevent over-fitting the training data (sufficiently complex networks can effectively memorise the training set, meaning that they can recover the parameters of the training set perfectly but have no predictive power otherwise), stopping the training when the loss function of the test/validation set stops improving.  

 Before inputting the spectra we normalise the data to have a mean of zero and standard deviation of 1 in each energy bin, by subtracting the mean count rate and dividing by the standard deviation. Similarly, we normalise the output parameters so that they are roughly similar in amplitude. In this case, it is sufficient to take the log of the column density and subtract the minimum value of $10^{20}$~cm$^{-2}$, while leaving the $\log(\xi)$ parameter covering the range from 1 to 3.
 We do not apply any rebinning to the data, and input the raw count rates in each channel between 0.5 and 5 keV directly into the network with no background or response matrix. These 450 channels correspond to the 450 input neurons in the network, and the two output neurons give the ionisation and column density parameter estimates. We experiment with various network architectures, and find that  a network with two hidden layers of 64 neurons performs well, with little or no improvement from making the layers larger or adding more layers.

We train the network for 5000 epochs with a batch size of 50, however the early stopping typically stops the training after $\sim1000$ steps. Once the network is trained, we apply it to the test sample, the same 1000 spectra that we fit in \textsc{xspec}. The parameter recovery is shown in the middle panels of Fig.~\ref{fig:parameter_estimates}. Relative to estimates from spectral fitting the neural network is slightly less accurate, with more scatter around the correct answer, but it also does not produce the outliers that occur with spectral fitting so the overall mean squared error is lower unless outliers are excluded (see Table~\ref{tab:methodcomparison}). The run time of the neural network is orders of magnitude faster than the spectral fitting approach, taking only 0.2 seconds to compute the parameter estimates for the 1000 spectra in the test set (this increases to $\sim0.5$~s when taking into account the time needed to load the model and spectra from files). 

\subsection{PCA preprocessing}
\label{sec:pca}

One reason that a neural net might struggle to accurately recover the model parameters is the complexity of the input dataset, as each unbinned spectrum has 450 energy bins, or 450 separate model inputs. In general, the performance of a machine learning model decreases with the number of inputs (or features) beyond some critical value. This is known as the `peaking phenomenon' or `Hughes phenomenon' \citep[][]{Hughes68,Trunk79}. A larger number of input dimensions requires a more complex network structure to process them, meaning that the network is more computationally expensive to train and needs a larger training set to achieve the same accuracy, and drastically increases the risk of overfitting the training set leading to poor performance on new data.

In reality, these energy bins are not independent, so the input data can be dramatically simplified, making the problem of parameter estimation correspondingly simpler. 
This is effectively a problem of dimensionality reduction: a 450 bin spectrum represents a single point in a 450 dimensional space, and the complete set of 11000 spectra some multi-dimensional shape. However, this shape could in principle be described much more efficiently with a different set of basis vectors as it will have negligible extent in most directions (analogous to a line or plane in 3D space). Principal Component Analysis (PCA) is a mathematical tool for reducing the dimensionality of datasets \citep[][]{Pearson1901}. It decomposes the data into a set of components (new basis vectors) and their amplitudes.  

Crucially, only the first few components will be needed to describe the dataset fully, with the higher order terms being attributable to noise. This means that each spectrum can be fully described by a much smaller number of parameters, simply by excluding these later components. This has an added benefit of removing much of the noise from the dataset. The number of components corresponding to real signal in the data can be identified with a log-eigenvalue (LEV) diagram, plotting the eigenvalues (the variance of the dataset along each of the new basis vectors) in order. The eigenvalues of components corresponding to noise follow a geometric progression (regardless of the source of the noise, as this is a property of the decomposition), which is generally quite distinct from the steeper decline seen in the first few components corresponding to genuine signal. While this method is somewhat subjective with noisier datasets (as it becomes harder to distinguish real components from noise when they have similar variance) this should not be a problem with training sets used for artificial neural networks, which should always be large enough for components to extracted well above the noise level.

Having established which components are genuine signals and which are due to noise, the noise components can be discarded, drastically reducing the number of parameters needed to fully describe the dataset. Dimensionality reduction like this is used extensively in machine learning applications, as it improves the performance of models with large input datasets, reduces noise in the data, allows algorithms that only work for low dimension inputs to be used, and reduces the resources needed to store and analyse the data.

\begin{figure}
    \centering
    \includegraphics[width=\linewidth]{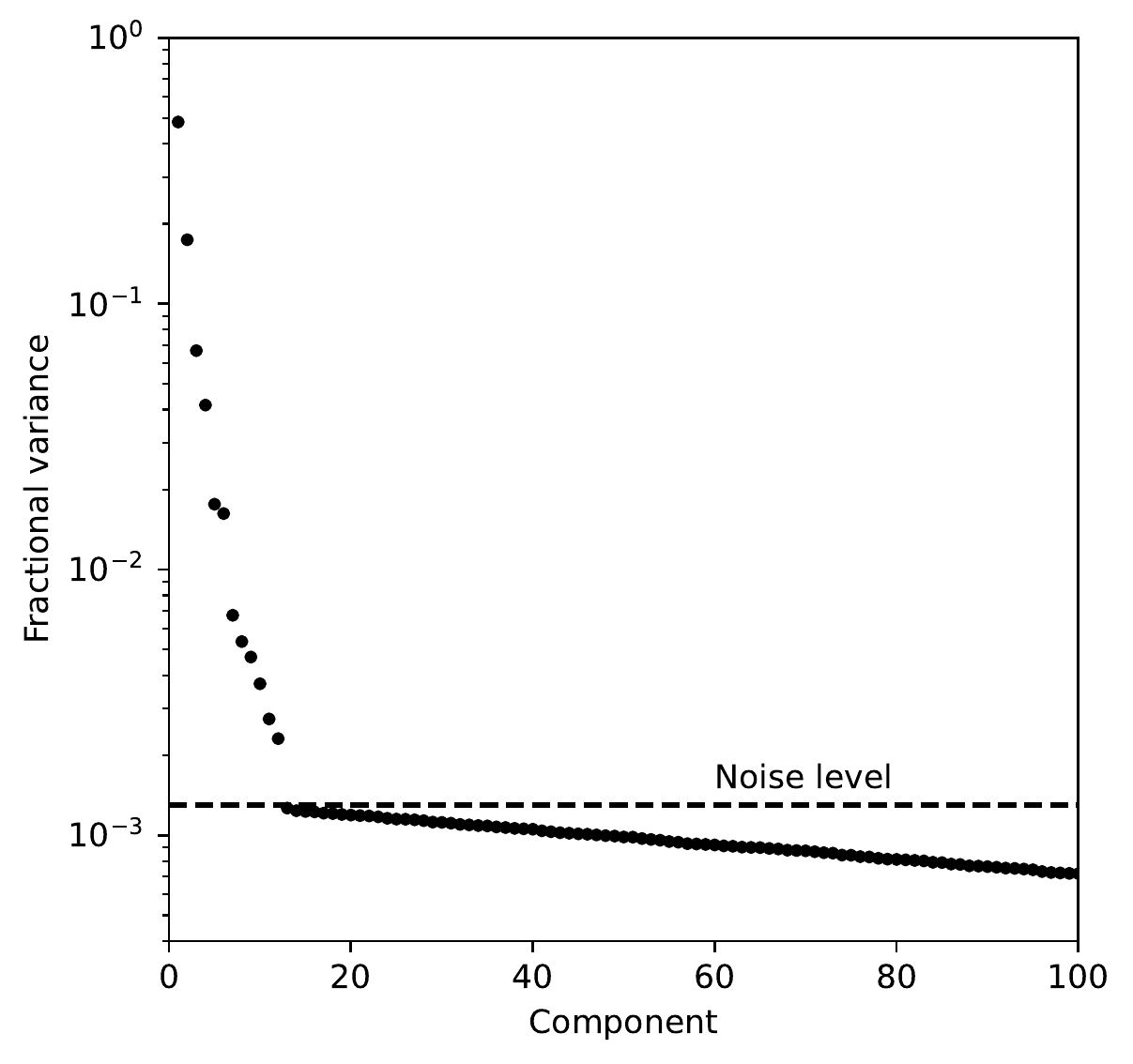}
    \caption{LEV diagram, showing the variance of the simulated dataset along each eigenvector. The first 12 correspond to real signal in the data, while the remaining components can be attributed to noise and excluded from the analysis.}
    \label{fig:lev}
\end{figure}

The shape of the principal components can also be informative, as they can correspond to the underlying physical components that make up the spectra \citep[e.g.][]{Parker14_mcg6}. However, for the purposes of this work, it is largely irrelevant whether these components have any particular physical meaning as we only require that the dimensionality of the data be reduced.

As in previous work \citep[][]{Parker15_pcasample}, we normalise the array of simulated spectra to fractional deviations from the mean count rate in each energy bin, before using Singular Value Decomposition (SVD) to perform the component analysis (using the \texttt{scipy.linalg.svd} function). The outputs from this function are the set of eigenvectors (the principal components), the corresponding eignevalues (the dispersion of the dataset along each eigenvector) and the coordinates of each spectrum in this new basis (analogous to the count rates in each energy bin).  

For our simulated dataset, the LEV diagram is shown in Fig.~\ref{fig:lev}. The first twelve components show significant variance, with the remainder attributable to noise. We therefore discard all components after the first twelve, meaning that the size of the full dataset has been reduced from $11000\times450$ to $11000\times12$ in the new basis. We then normalise the new dataset to have a mean of zero and standard deviation of 1 in each column, and input it into the same neural network architecture. This reduction will be even more effective for higher resolution spectra, such as those of the \athena\ X-ray Integral Field Unit (XIFU), which will have thousands of energy bins.

\begin{table}
    \centering
    \caption{Performance of the different parameter estimation techniques. For the spectral fitting, we show the mean squared error for all data points, and with outliers (defined as points with an individual squared error of $>1$) excluded. With outliers excluded spectral fitting is the most accurate, otherwise the neural net with PCA pre-processing is the most accurate.}
    \begin{tabular}{l c r}
    \hline
    \hline
    Method & Run Time (s) & Mean squared error\\
    \hline
    Spectral fitting & 330 & $0.12$ ($2.6\times10^{-4}$) \\
    Neural network     & 0.2 & $4.2\times10^{-3}$\\
    PCA $+$ Neural network & 0.4 & $7.0\times10^{-4}$\\
    \hline
    \end{tabular}
    \label{tab:methodcomparison}
\end{table}

We then input the compressed dataset into a neural network, as in Section~\ref{sec:neuralnet}. The dimensionality reduction means that the input layer is only 12 neurons rather than 450, and we find that two hidden layers of 32 neurons is sufficient for this network, with wider or deeper networks offering no significant improvement. We show a comparison of the architecture of this network and the one from Section~\ref{sec:neuralnet} in Table~\ref{tab:networks}. The number of  free parameters of the network (i.e. the total number of weights and biases of the neurons) is drastically reduced by the PCA step, with 1,538 parameters in this model compared to 33,154 in the pure neural network model.

\begin{figure*}
    \centering
    \includegraphics[width=0.7\linewidth]{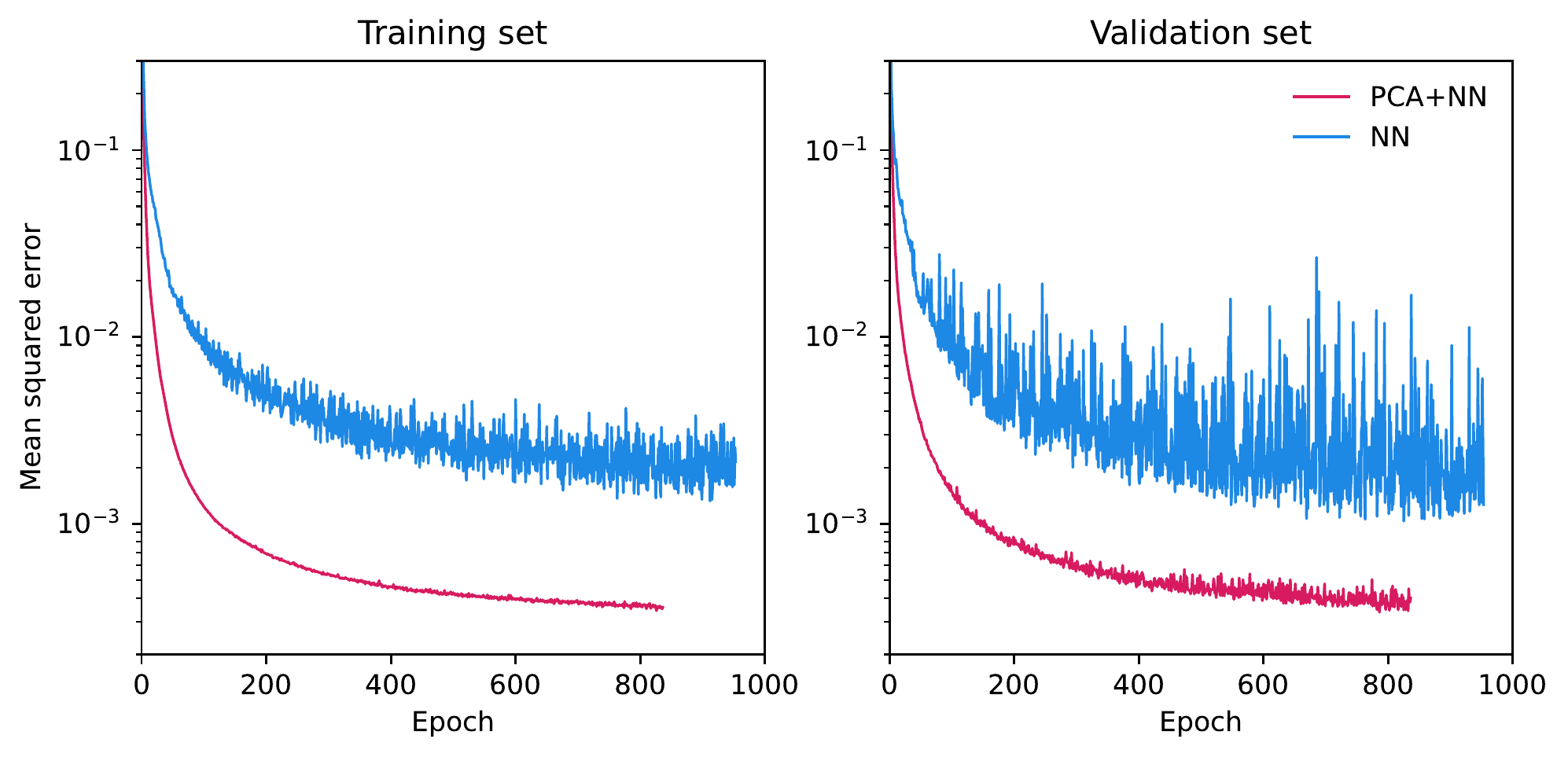}
    \caption{Training history of the neural network with and without PCA preprocessing. The PCA version converges to a lower mean square error solution faster and with much less noise in both the training and validation set. Note that the mean squared error shown here is not directly comparable to that in Table~\ref{tab:methodcomparison} as these values are calculated from the normalised parameters used by the networks.}
    \label{fig:history}
\end{figure*}

The training performance of the network is significantly better after the PCA step. We show the training history (i.e. the performance of the neural net as a function of training epoch) of the networks with and without PCA in Fig.~\ref{fig:history}. The network with PCA preprocessing converges faster on a better solution, and with far less noise.
The resulting parameter estimation is shown in the bottom panels of Fig.~\ref{fig:parameter_estimates}. The parameter recovery is very good, with no outliers and much smaller scatter than the pure neural network approach (a factor of 2 reduction in the standard deviation of the error in $\log \xi$ and a factor of 3 in $\log N_\mathrm{H}$). The additional computational expense is very small, with the PCA taking 0.2~s to reduce the dimensionality of the dataset. 

\begin{table}
    \centering
    \caption{Architecture of the two neural networks used in this work.}
    \begin{tabular}{c c}
    \hline
    \hline
         Pure NN & PCA + NN \\
         \hline
                &   Normalise spectra \\
                &   PCA (450 channels to 12 eigenvectors) \\
            Normalise spectra    &   Normalise amplitudes \\
            Input layer (450 neurons)    &  Input layer (12 neurons) \\
            Hidden layer (64 neurons)    &  Hidden layer (32 neurons)\\
            Hidden layer (64 neurons)    &  Hidden layer (32 neurons)\\
            Output layer (2 neurons)     &  Output layer (2 neurons)\\
    \hline
    \end{tabular}
    \label{tab:networks}
\end{table}

A secondary benefit of the PCA approach is that it can quickly identify spectra that are outside the assumptions used to train the network. Any input spectrum can be trivially transformed into the new basis, and spectra that are consistent with those in the training set should be well described purely by the first twelve components. If the contribution from higher order components increases beyond what would be expected for noise it implies that the decomposition is not valid for this new spectrum, and that it is likely not consistent with the assumptions of the training set (for example, because it has additional spectral components in it).

\section{Discussion}
\label{sec:discussion}
 
We have explored the use of neural networks for parameter estimation in X-ray spectra, looking specifically at warm absorbers in \athena\ WFI spectra. Overall, our results are promising, demonstrating that a neural network can in principle deliver similar accuracy to a conventional spectral fitting approach in a fraction of the time, and without the problem of false minima that frequently affects automated spectral fitting. We have further shown that pre-processing the spectra with PCA increases the accuracy of the parameter estimation.  
 
 \subsection{Comparison with other work}
While neural networks have been used in data analysis for a long time, their use in X-ray astronomy to date has been minimal, due to the relatively small quantities of data produced by X-ray instruments. This is guaranteed to change with future instrumentation, as the volume of data will be orders of magnitude larger than it currently is.
\citet[][]{Ichinohe18} demonstrated that a neural network can be used to estimate physical parameters directly from an X-ray spectrum, considering a single temperature thermal plasma in simulated \emph{Hitomi} spectra. We build on this to show that a similar approach can be used on lower resolution, shorter exposure data with additional confusing components, of the type that is likely to benefit most from a machine-learning approach.

\citeauthor{Ichinohe18} find that more complex networks are required to maximise the performance, but achieve a similar level of accuracy to this work. This difference may be due to the higher number of energy bins in the microcalorimeter spectra they considered (7200, compared to 450). The importance of using pre-processing with PCA or other dimensionality reduction tools to reduce the complexity of the input data is likely to increase with the number of energy bins, as reducing the number of inputs should also reduce the complexity of the network needed to analyse them.
 
 In other branches of astronomy where the datasets are larger, neural network based approaches are more common. For example, some other authors have considered the use of PCA pre-processing and neural networks for classifying stellar spectra \citep[e.g.][]{StorrieLombardi94, Singh98} or galaxy spectra \citep[][]{Folkes96}, generally finding that the PCA pre-processing allows for simpler network architectures and improved classification accuracy.  

Neural networks can also be used for solving the inverse problem of predicting spectra from parameters, approximating much more complex models \citep[e.g.][]{Alsing20}. By training the network on a representative sample of model spectra, it can predict spectra for other parameter combinations, meaning that larger and more accurate (relative to interpolation between grid points) models can be generated cheaply. We explore this for modelling Ultra-fast outflow (UFO) spectra in AGN in Matzeu et al. (submitted), finding that spectra can be generated more accurately than through interpolation and in a tiny fraction of the time needed to run the full model code.

\subsection{Training sets}
The accuracy of the neural networks is dependent on how reliable and how representative the training set is. In this case, the training set is synthetic spectra covering the same parameter space as the test set, and both are simulated using the same model, so the training set meets both criteria. When applying this technique to real data, the training set will need to either be synthetic spectra calculated using some assumed model, or a smaller subsample of real spectra with parameters estimated by spectral fitting. In each case, the reliability of the final parameter estimates will depend on the reliability of the spectral models used. It may be worth training multiple networks on slightly different training sets to mitigate the possibility of model dependence. For example, networks could be trained on synthetic spectra simulated with different models, and on a subset of real spectra that have been manually analysed.

\subsection{Classification}

The neural nets we have trained here rely on the assumptions of a fairly simple continuum and a single layer warm absorber. If these assumptions are not met, then the parameters returned are likely to be systematically biased or meaningless, as the parameters from spectral fitting would be if the incorrect model was used. For example, if no warm absorber is present then the network will still return estimates of ionisation and column density. It is therefore essential that the spectra first be classified to ensure that the appropriate neural networks are applied to it. This also means that the neural networks used for parameter estimation can be smaller and more focused on spectra in a particular regime, rather than trying to train a single huge network to estimate all parameters in all possible spectra. Given the variety and complexity of X-ray AGN spectra, it is probably unrealistic to construct a single parameter estimation tool that can be applied universally, in the same way that it would be unwise to construct an \textsc{xspec} model with every possible physical component in it simultaneously.

Some simple cuts can be made on hardness ratios, which should rule out strongly absorbed or jet-dominated sources, for example. A machine-learning approach could also be used, for example by training neural networks {or random forest classifiers} to identify spectra {of AGN from particular classes or} with particular features, such Compton thick AGN, or AGN with warm absorbers, soft excesses, or broad iron lines. Those spectra could then be passed to appropriate parameter estimation tools. This stage would be essentially equivalent to the step in spectral fitting where the user inspects the spectrum visually and decides which models to apply. {While identifying the most efficient solution to this classification problem is far outside the scope of this work, we speculate that a three layer system with the first classifier to select a general class of object (type I AGN), then a second to select sources of interest within that class (those with warm absorbers), and parameter estimation done by the third layer. We note that such classification problems have been studied extensively in other branches of astronomy \citep[e.g.][]{Eatough10,Dieleman15,Kim16,Aniyan17,Osborn20}.}

\subsection{Error estimation}
We have not evaluated the errors on the parameter estimates for any of the methods explored. In general, it is obviously very important to know the uncertainty in parameter estimation, along with possible systematics. For spectral fitting, the conventional procedure is either to use an error estimation algorithm in the fitting package that calculates the error by adjusting parameters and refitting until the fit statistic crosses some threshold, or by using Monte Carlo Markov Chains (MCMC) to map the parameter space. Both of these methods are very computationally expensive on large scales. Running the \textsc{xspec} error calculations on a single spectrum with out simple model takes around 5~seconds, which would increase the total time required to fit the sample by over an order of magnitude, and a robust MCMC would typically take several minutes per spectrum, even with a small number of parameters.

The fastest and simplest way to estimate the uncertainties in the parameter estimates for all three methods is simply to quantify the scatter in their estimates of parameters in the test set. A more sophisticated approach for the neural network based approaches may be to use a Bayesian neural network, which outputs probability distributions instead of single values for parameters. This is likely to perform better with more complex model parameter spaces, where the uncertainty on any given parameter depends on the values of multiple other parameters. A Monte-Carlo approach can also be used by adding dropout layers to the model after each layer \citep{Gal2016dropout}. These layers will randomly zero a set fraction of the inputs to the next layer at each step, which is used during training to prevent overfitting. However, if they are used during the test phase as well, then the parameter estimates will be slightly different each time, and the resulting scatter gives an estimate of the uncertainty. {This increases the run time by a factor of the number of iterations, but this is still likely to be very small compared to the time needed for conventional error estimation. More problematic for this approach is the need to have a more complex network to achieve the same accuracy, to compensate for the information lost by the dropout layers.}

\subsection{Advantages and disadvantages of the machine learning approach}
{While neural networks can be applied to almost any problem, this does not mean that they should be. There are many problems in astronomy where there is no real advantage to using a neural network, where an answer can be obtained more quickly and more accurately with existing techniques. A realistic evaluation of whether machine learning is ultimately useful for a given problem is therefore essential (and ideally should be undertaken in advance).}

{The main advantages of neural network are the speed, and the scalability, as we have discussed throughout. The main computational cost involved in setting up a network is in the training process, which is essentially fixed regardless of the size of the dataset to be evaluated. A rough threshold for when a neural network will become useful for spectroscopy is when the time taken to train the network is small relative to the time needed to model the spectra conventionally. In this work this condition is not met, as we only consider a test set of 1000 spectra which take 5 minutes to fit conventionally, while the network takes $\sim10$ minutes to train, but the fitting time scales linearly with the number of spectra, while the training time is fixed.}

A related issue is that the neural network is only valid for the specific data on which it was trained, and cannot trivially be applied to data from a different instrument. For example, if a new instrument response is released then the network will need to be re-trained to account for this. {In the analogous situation for conventional fitting, the spectra would have to be re-fit using the updated response files}, but these fits could be initialised from the previous best fit and would likely be faster to converge {and no changes to the fitting algorithm would be needed}. 
In fact, a similar approach could be taken with a neural network, using transfer learning \citep[e.g.][]{zhuang2020comprehensive} to update the weights of the network without starting from scratch. This would be much faster, and require less training data. An additional interesting possibility is that for minor changes in the model or instrument response it might be possible to update the PCA decomposition without retraining the network, which would still be valid if the new PCA components correspond to the old ones. This would remove the need for computationally expensive retraining, but would need to be carefully tested to ensure the validity of the network.

{For a small number of spectra, such as a handful of observations of a single source, the machine learning approach is clearly worse than conventional spectroscopy performed manually, as it will be slower overall, less accurate, and far less flexible than an informed user carefully modelling the spectrum. There may still be some use to neural networks in this regime, for example obtaining rough parameter estimates to initialise fits, however the computational expense of training a suitable network likely outweighs the time saved.}

{In this work, we have considered survey data specifically, where large numbers of spectra of moderate quality need parameter estimation from relatively simple models. In this regime, machine learning is extremely useful, as it takes a very small fraction of the time of automated spectral fitting, and avoids the problem of false minima. This conclusion is likely to hold in any case where very large numbers of spectra need to be analysed, such as large scale surveys or very high time resolution spectroscopy of variable sources. The more computationally expensive the spectral model is to evaluate, the greater the speed advantage of the neural network. It is also likely that neural networks will outperform automated spectral fitting in complex parameter spaces with false minima, as the parameter space we considered here was relatively simple and the automated fits still get stuck occasionally. The PCA pre-processing also conveys an advantage with higher resolution data, as in general it is more computationally expensive to evaluate fits on spectra with more energy bins. PCA sidesteps this issue by reducing the dimensionality of the data input to the neural network, so the network can remain small regardless of the number of bins of the original spectra.}
{A related hidden cost of the neural net training process is the time needed to evaluate and decide on a network architecture, as there is no fixed solution for this. PCA helps here as well, as it allows a smaller network to be used, making it faster to train and more generally applicable.}

\section{Conclusions}
\label{sec:conclusions}

We have explored how artificial neural networks can be used to estimate physical parameters from raw X-ray spectra, without needing to go through a computationally expensive and potentially unreliable automated spectral fitting process. We use simulated \athena\ WFI spectra to train the networks, and compare their ability to recover the parameters of a test set of spectra with a conventional spectral fitting approach. Our main findings are:

\begin{itemize}
    \item Relatively simple neural networks with two hidden layers and 32--64 neurons in each layer are able to reliably recover the ionisation and column density of a warm absorber in synthetic spectra. This approach is orders of magnitude faster than automated spectral fitting, allowing huge volumes of spectra to be processed in seconds. 
    \item For most spectra, the spectral fitting approach is more accurate, however for a minority of spectra the fit gets stuck in a false minimum, missing the correct solution by a wide margin. If these outliers are taken into account, then the neural network approach is more accurate as well as being faster. If the outliers are excluded then the accuracy of the best performing network is slightly worse than that of spectral fitting (a factor of 2--3 higher in mean squared error).
    \item Including a pre-processing step using PCA to reduce the dimensionality of the data prior to input into the neural network means that higher accuracy (a factor of 6 reduction in mean squared error) can be achieved with a simpler network (32 rather than 64 neurons in the hidden layers, and an input layer of 12 rather than 450 neurons). This has the added benefit of removing noise from the dataset, and adds a mechanism for identifying when new spectra are outside the assumptions of the training set. 
\end{itemize}

We conclude that this is a promising approach for the large-scale analysis of X-ray data from the next generation of instrumentation, such as \athena , allowing rapid analysis without compromising on accuracy. Use cases include large scale surveys or studies of individual objects at high time resolution {where the number of spectra to be evaluated is very large}. {We anticipate little use for this kind of approach with individual or small numbers of spectra, as the training process is more computationally expensive than spectral fitting in this regime.}

\section*{Acknowledgements}
ML acknowledges a Machine Learning $\&$ Cosmology Research Fellowship from the University of Nottingham. GAM acknowledges the financial support from Attività di Studio per la comunità scientifica di Astrofisica delle Alte Energie e Fisica Astroparticellare: Accordo Attuativo ASI-INAF n. 2017-14-H.0.

\section*{Data availability}
The response and background files used to simulate the synthetic spectra are available from the WFI website: \url{https://www.mpe.mpg.de/ATHENA-WFI/response_matrices.html}. All software packages used are publicly available. The analysis scripts used in this analysis can be provided on reasonable request to the authors.




\bibliographystyle{mnras}
\bibliography{bibliography} 







\bsp	
\label{lastpage}
\end{document}